\documentclass{optica-article}

\journal{opticajournal} 

\articletype{Research Article}
\usepackage[T1]{fontenc}
\usepackage{caption}
\usepackage{listings}
\usepackage{lineno}
\usepackage[version=4]{mhchem}
\usepackage{subcaption}
\usepackage{graphicx} 
\usepackage{amsmath}
\usepackage{siunitx}
\usepackage{physics}
\usepackage[capitalise]{cleveref}
\graphicspath{{./Figures/}}
\begin{document}

\title{Stability of a high-finesse optical cavity at 493~nm in vacuum for cavity QED with Barium ions}

\author{Diptaranjan Das\authormark{1*}, Ezra Kassa,\authormark{1} and Hiroki Takahashi\authormark{1+}}

\address{\authormark{1} Experimental Quantum Information Physics Unit, 
Okinawa Institute of Science and Technology, 
1919-1 Tancha, Onna-son, Okinawa, 904-0495, Japan\\
}

\email{\authormark{*}diptaranjan.das@oist.jp} 
\email{\authormark{+}hiroki.takahashi@oist.jp}



\begin{abstract*} 
We explore the stability of a high-finesse optical cavity at 493~nm in vacuum for cavity QED with Barium ions. 
A high-finesse Fabry-Perot cavity is built using mirrors with high-reflectivity (HR) coatings that are implemented by stacking multiple thin films of low-loss dielectrics on substrates. 
Applications of such HR mirrors in the near ultraviolet (UV) range have been hampered by degradation of coatings in vacuum. 
Here, we explore the degradation of mirrors with HR coatings at 493~nm in vacuum. 
We study both vacuum-induced and laser-induced effects on oxide-coated cavity mirrors by probing changes in cavity loss using cavity lifetime measurements. 
We investigate the role of circulating power in the rate of increase in cavity loss and demonstrate methods of reversal of cavity degradation. 
While we observe no degradation without long exposure or with short exposures at lower circulating powers, we find evidence of degradation on long exposure to high circulating powers. 
We discuss potential causes and conclude that laser-induced deposition is the likely cause while ruling out thermally activated processes due to laser-induced heating.  
\end{abstract*}

\section{Introduction}

The confinement of light in high-finesse optical cavities has opened up avenues for strong light-matter interaction. 
In recent years, there has been a proliferation in research activity in this area, with smaller mode volumes in optical cavities and highly reflecting mirror surfaces making it possible to achieve strong coupling between light and matter \cite{mabuchi2002cavity, Kimble1997, HJKimble1998}. 
To this end, high-reflectivity (HR) optical coatings with low loss and defects have played an important role. They have been used to build high-finesse cavities for coupling to trapped ions, atoms and solid state emitters \cite{HJKimble1998, JYe1999, McKeever2003,Takahashi2020}. 
Apart from the interests in fundamental physics in cavity quantum electrodynamics (QED) with strong coupling \cite{Kimble1997, HJKimble1998, McKeever2003,Takahashi2020}, there are several applications of these systems, such as efficient single photon sources \cite{Keller2004, Wilk2007}, implementing nodes in a quantum network for quantum communication \cite{Kimble2008,Ritter2012}, exploring non-linear optics at a quantum level \cite{Chang2014} and building quantum simulators for exploring exotic physics \cite{reitz2022cooperative}. 
 
\begin{figure}[ht]
    \centering 
    \noindent
    \begin{subfigure}[t]{0.4\textwidth}
        \centering
        \def\svgwidth{\linewidth}
\begingroup%
  \makeatletter%
  \providecommand\color[2][]{%
    \errmessage{(Inkscape) Color is used for the text in Inkscape, but the package 'color.sty' is not loaded}%
    \renewcommand\color[2][]{}%
  }%
  \providecommand\transparent[1]{%
    \errmessage{(Inkscape) Transparency is used (non-zero) for the text in Inkscape, but the package 'transparent.sty' is not loaded}%
    \renewcommand\transparent[1]{}%
  }%
  \providecommand\rotatebox[2]{#2}%
  \newcommand*\fsize{\dimexpr\f@size pt\relax}%
  \newcommand*\lineheight[1]{\fontsize{\fsize}{#1\fsize}\selectfont}%
  \ifx\svgwidth\undefined%
    \setlength{\unitlength}{742.5bp}%
    \ifx\svgscale\undefined%
      \relax%
    \else%
      \setlength{\unitlength}{\unitlength * \real{\svgscale}}%
    \fi%
  \else%
    \setlength{\unitlength}{\svgwidth}%
  \fi%
  \global\let\svgwidth\undefined%
  \global\let\svgscale\undefined%
  \makeatother%
  \begin{picture}(1,0.58080808)%
    \lineheight{1}%
    \setlength\tabcolsep{0pt}%
    \put(0,0){\includegraphics[width=\unitlength,page=1]{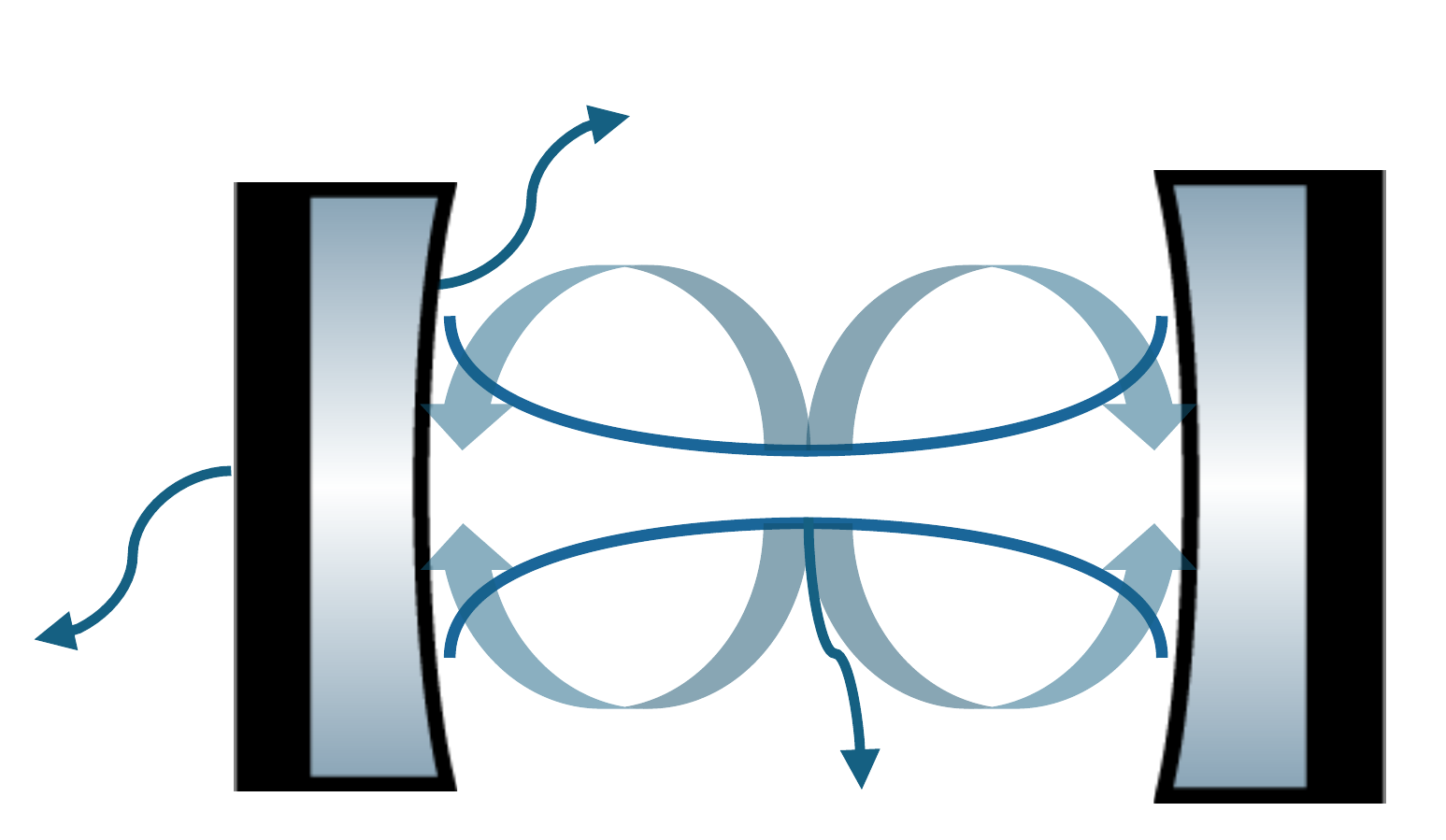}}%
    \put(0.54192116,0.39982457){\color[rgb]{0,0,0}\makebox(0,0)[lt]{\lineheight{1.25}\smash{\begin{tabular}[t]{l}$g$\end{tabular}}}}%
    \put(0.44555948,0.49099291){\color[rgb]{0,0,0}\makebox(0,0)[lt]{\lineheight{1.25}\smash{\begin{tabular}[t]{l}$\kappa_{in}$\end{tabular}}}}%
    \put(-0.04175408,0.09776226){\color[rgb]{0,0,0}\makebox(0,0)[lt]{\lineheight{1.25}\smash{\begin{tabular}[t]{l}$\kappa_{ex}$\end{tabular}}}}%
    \put(0.61209143,0.02287215){\color[rgb]{0,0,0}\makebox(0,0)[lt]{\lineheight{1.25}\smash{\begin{tabular}[t]{l}$\Gamma$\end{tabular}}}}%
    \put(0,0){\includegraphics[width=\unitlength,page=2]{cavityqed.pdf}}%
  \end{picture}%
\endgroup%

        \caption{}
        \label{figschematicwithatoms}
    \end{subfigure}
    \hspace{1cm}
    \begin{subfigure}[t]{0.4\textwidth}
        \centering
        \def\svgwidth{\linewidth}
\begingroup%
  \makeatletter%
  \providecommand\color[2][]{%
    \errmessage{(Inkscape) Color is used for the text in Inkscape, but the package 'color.sty' is not loaded}%
    \renewcommand\color[2][]{}%
  }%
  \providecommand\transparent[1]{%
    \errmessage{(Inkscape) Transparency is used (non-zero) for the text in Inkscape, but the package 'transparent.sty' is not loaded}%
    \renewcommand\transparent[1]{}%
  }%
  \providecommand\rotatebox[2]{#2}%
  \newcommand*\fsize{\dimexpr\f@size pt\relax}%
  \newcommand*\lineheight[1]{\fontsize{\fsize}{#1\fsize}\selectfont}%
  \ifx\svgwidth\undefined%
    \setlength{\unitlength}{170.79874474bp}%
    \ifx\svgscale\undefined%
      \relax%
    \else%
      \setlength{\unitlength}{\unitlength * \real{\svgscale}}%
    \fi%
  \else%
    \setlength{\unitlength}{\svgwidth}%
  \fi%
  \global\let\svgwidth\undefined%
  \global\let\svgscale\undefined%
  \makeatother%
  \begin{picture}(1,0.66334408)%
    \lineheight{1}%
    \setlength\tabcolsep{0pt}%
    \put(0,0){\includegraphics[width=\unitlength,page=1]{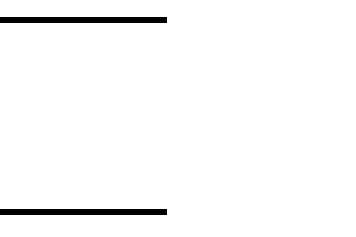}}%
    \put(0.47998408,0.02963319){\color[rgb]{0,0,0}\makebox(0,0)[lt]{\lineheight{1.25}\smash{\begin{tabular}[t]{l}$S$ or $D$\end{tabular}}}}%
    \put(0.47658786,0.56974098){\color[rgb]{0,0,0}\makebox(0,0)[lt]{\lineheight{1.25}\smash{\begin{tabular}[t]{l}$P$\end{tabular}}}}%
    \put(0.10615021,0.2304354){\color[rgb]{0,0,0}\makebox(0,0)[lt]{\lineheight{1.25}\smash{\begin{tabular}[t]{l}$p\Gamma$\end{tabular}}}}%
    \put(0.6283279,0.43442844){\color[rgb]{0,0,0}\makebox(0,0)[lt]{\lineheight{1.25}\smash{\begin{tabular}[t]{l}$(1-p)\Gamma$\end{tabular}}}}%
    \put(0,0){\includegraphics[width=\unitlength,page=2]{level_diagram.pdf}}%
  \end{picture}%
\endgroup%

        \caption{}
        \label{figleveldiagram}
    \end{subfigure}
    \\
    \vspace{1cm}
    \noindent
    \begin{subfigure}[t]{0.7\textwidth}
        \begin{tabular}{|c c c c c|} 
        \hline
        Species & Transition & $\lambda$ (\unit{\micro\meter}) & Branching  & $p\lambda$ (\unit{\micro\meter})\\ 
        &  &  &  ratio ($p$) & \\[0.5ex] 
        \hline
        Ca\textsuperscript{+} &  $4{}^2S_{1/2}-4{}^2P_{1/2}$ & 0.397 &   0.936  & 0.372\\
        \hline
        Ca\textsuperscript{+} & $3{}^2D_{3/2}-4{}^2P_{1/2}$ & 0.866 & 0.064   & 0.051\\
        \hline
        Ca\textsuperscript{+} & $3{}^2D_{5/2}-4{}^2P_{3/2}$ & 0.854 & 0.059   &  0.050\\
        \hline
        
        Sr\textsuperscript{+} &  $5{}^2S_{1/2}-5{}^2P_{1/2}$ & 0.422 & 0.941 &  0.397 \\
        \hline
        Sr\textsuperscript{+} & $4{}^2D_{3/2}-5{}^2P_{1/2}$ & 1.091 &  0.055  & 0.060\\
        \hline
        Sr\textsuperscript{+} & $4{}^2D_{5/2}-5{}^2P_{3/2}$ & 1.033 &  0.053  &  0.055\\
        \hline
        
        Ba\textsuperscript{+} &  $6{}^2S_{1/2}-6{}^2P_{1/2}$ & 0.493 &   0.729  & 0.359\\ 
        \hline
        Ba\textsuperscript{+} & $5{}^2D_{3/2}-6{}^2P_{1/2}$ & 0.650  &   0.271  & 0.176\\
        \hline
        Ba\textsuperscript{+} & $5{}^2D_{5/2}-6{}^2P_{3/2}$ & 0.614  &  0.215  & 0.132\\
        \hline
        
        Yb\textsuperscript{+} &  $6{}^2S_{1/2}-6{}^2P_{1/2}$ & 0.370 &  0.995  & 0.368\\
        \hline
        Yb\textsuperscript{+} &  $5{}^2D_{3/2}-5d6s{}^3D[3/2]_{1/2}$ & 0.935 & 0.017 &  0.016\\
        \hline 
        \end{tabular}
        \caption{}
        \label{table1}
    \end{subfigure}
    \caption{ (\subref{figschematicwithatoms}) Model for cavity QED with a single atom: $g$ is the coherent coupling of the atom to the cavity field. $\Gamma$ is the incoherent spontaneous emission rate of the atom into free space. 
    The cavity decay rate $\kappa$ is the sum of the external coupling rate $\kappa_{\mathrm{ex}}$ and the internal loss of the cavity field $\kappa_{\mathrm{in}}$. 
    (\subref{figleveldiagram}) A simplified level diagram of the ion. The ion can spontaneously decay to the lower $S$ or $D$ level from the excited $P$ level at a rate given by $\Gamma$ and the branching ratio $p$. 
    (\subref{table1}) Table showing transition wavelengths ($\lambda$) of different ionic species, branching ratios ($p$) and their products ($p\lambda$).}
\end{figure}

A model for cavity QED is shown in \cref{figschematicwithatoms}. When the resonance of the cavity is tuned to an atomic transition, the atom located at the antinode of the cavity field is coupled to and coherently exchange energies with the cavity mode at a rate $2g$.
Strong coupling is then determined by the condition $C=\frac{g^{2}}{\Gamma \kappa}>1$, where $\Gamma$ is the incoherent spontaneous emission of the atom into free space, $\kappa$ is the amplitude damping rate of the cavity field and $C$ is known as the cooperativity parameter.
In quantum communication, rapid generation of entanglement can be driven by the strong coupling of light with atoms, while efficient collection of photons can be made possible by the highly directional nature of photons transmitted by the cavity. 

Among various systems that can be interfaced with optical cavities, trapped ions possess remarkable features such as long trapping lifetimes, long coherence times \cite{Wang2021} and high-fidelity quantum operations \cite{Harty:14}. 
As a result, trapped ions have been the subject of intense research in cavity QED in recent decades \cite{Keller2004,Stute2012,Takahashi2020}. 
Many major ionic species such as Ca\textsuperscript{+}, Sr\textsuperscript{+} and Ba\textsuperscript{+}  share a common energy level structure in which there are the ground ($S$), excited ($P$) and metastable ($D$) levels. 
The cavity can be tuned to be resonant to either the $S$-$P$ or $D$-$P$ transition. 
The excited $P$ state can decay to the $S$ or $D$ states at rates given by the spontaneous emission rate $\Gamma$ with a branching ratio $p$ that depends on the atomic species (see \cref{figleveldiagram}). 
Assuming that the ion is coupled to the TEM\textsubscript{00} mode of the cavity, the cooperativity parameter can also be written as follows \cite{kassa2024integrate}:
\begin{align}
    C &= \frac{6F}{\pi^2}\frac{p\lambda}{\sqrt{l\qty(2R_c-l)}}, \label{cooperativity}
\end{align}
where $F$ is the finesse of the cavity, $\lambda$ the resonance wavelength, $l$ the cavity length, $R_c$ the radius of curvature of the mirrors. As seen in \cref{cooperativity}, for a given cavity geometry and finesse, the atomic properties of the ion only enter as a product $p\lambda$.
As shown in the table of \cref{table1}, the $S$-$P$ transition lies at a visible or ultra-violet (UV) wavelength. 
On the other hand, the $D$-$P$ transitions are in the near-infrared (NIR) regime. 
In general, using the $S$-$P$ transition is beneficial in achieving a large cooperativity parameter due to the high branching ratios. As can be seen in \cref{table1} \footnote{The $D_{3/2}-D[3/2]_{1/2}$ transition of Yb\textsuperscript{+} is an exception in \cref{table1} that does not have a P-level as an excited state.}, the $S$-$P$ transitions exhibit higher $p\lambda$ (>0.35 \unit{\micro\meter}) than other transitions.
However, so far most experiments in cavity QED with trapped ions have exploited optical cavities resonant with the NIR transitions \cite{Keller2004,Stute2012,Steiner2013,Takahashi2020}. 
This is because cavity mirror coatings at UV wavelengths often degrade in vacuum, resulting in an increase in mirror loss and a drop in cavity finesse. In both\cite{Sterk2012} and \cite{Ballance2017} where Yb\textsuperscript{+} ions were coupled to cavities resonant at a wavelength of $369$~nm, degradation of the cavity mirrors was observed.

So far, there have been several hypotheses for the underlying mechanisms leading to such damage. The state-of-the-art highly reflecting coatings are implemented by stacking multiple thin films of dielectric oxides, typically Silica and Tantalum (V) oxides, on substrates using a method such as ion beam sputtering. 
Oxygen diffusion from the top oxide layer in vacuum has been discussed as a possible mechanism to incur damage to the coating \cite{Gangloff2015}. 
The process creates absorption centers on the top surface and is not laser induced. In \cite{Gangloff2015}, it has been argued that the silica top layer is more robust to this type of damage, although there are disagreements \cite{Ballance2017, Gallego}. 
On the other hand, the deposition of outgassing hydrocarbons on coated surface in vacuum on exposure to UV radiation has also been suggested and experimentally investigated under various conditions \cite{Schmitz2019,wagner2014laser}. In \cite{Schmitz2019}, the authors observe no degradation to their cavity in vacuum on exposure to UV radiation in the absence of hydrocarbon materials in the setup. The cavity was kept in vacuum for several months. 
However, it is unclear how long the cavity was exposed to radiation in total, since the finesse is measured by scanning the cavity. 
The duration of exposure is therefore likely to be short, with the cavity being exposed to resonant light for only a short duration during the scan to measure the finesse. 
This may be a vital factor, since the processes driving degradation are likely to depend on the cumulative intensity of radiation that the mirror is exposed to. 
Other experiments have used high intensity pulsed lasers of short wavelength and measured surface reflectivity to test for degradation \cite{wagner2014laser}. 
Some experiments have indicated that the micro-chemical environment of the top layer may play a factor \cite{brand2023multi}. 
Damage, irrespective of speculated mechanisms, has been shown to be reversible, partially or fully, by exposure of the mirrors to oxygen \cite{Gangloff2015,brand2023multi,Ballance2017}. 
The presence of radiation at $422$~nm along with exposure to oxygen was also shown to accelerate recovery \cite{Gangloff2015}. An increase in surface roughness leading to an irreversible drop in finesse was observed when oxide coated mirrors were baked at a temperature of 180$^{\circ}$ under ultra-high vacuum \cite{Rudelis}.
Damage is typically observed at short wavelengths \cite{Gangloff2015, Ballance2017}, and is known to occur deep in the ultraviolet zone \cite{wagner2014laser} as well. It is not known whether similar degradation of cavity mirrors in vacuum would occur in the visible range near $500$~nm. Experiments involving $554$~nm laser have not shown any damage to high finesse cavities \cite{Takei2010}. 
Investigations have also been carried out for infrared radiation, showing no such damage in some instances \cite{Takahashi2020}, while damage has been reported at $770$ \unit{\nano\meter} \cite{Gallego}. 
Degradation of cavity mirrors, designed to operate at $854$~nm, on annealing in vacuum has been observed in \cite{Brandstatter2013}. 

In this paper, we investigate the feasibility of using mirrors that are coated for high reflectivity at $493$~nm in vacuum for the implementation of cavity QED using the $S_{1/2}$-$P_{1/2}$ transition of Ba\textsuperscript{+} ions. The wavelength of this transition is furthest from the UV band among the equivalent transitions in commonly used ionic species (see Table in \cref{table1}). 
Strong coupling in Calcium ions has been demonstrated, with the cavity being coupled to the $P_{1/2}$-$D_{3/2}$ transition at $866$~nm \cite{Takahashi2020}. 
Stronger coupling can be achieved using the $S_{1/2}$-$P_{1/2}$ transition of Calcium at $397$~nm but any efforts have been hampered by cavity degradation at that wavelength \cite{KellerPrivate}. 
Cavity coupling using the $S_{1/2}$-$P_{1/2}$ transition in Barium ion can reach a similarly high cooperativity at $493$~nm, which is approximately $100$~nm longer than Calcium's. 
As an example, a cavity length of $500$~\unit{\micro\meter} using a pair of mirrors with a radius of curvature $450$~\unit{\micro\meter} can lead to coupling strength of $2\pi\times40$~MHz. 
For excited state spontaneous emission rates (for $\text{Ba}^+$, decay rate from $P_{1/2}$ level, $\Gamma=2\pi\times20.4$ MHz) and a realistic cavity decay rate of $\kappa=2\pi\times4$ MHz \cite{Takahashi2020}, we expect cooperativity $C\approx 20$, being in the strong coupling regime.
Even though the critical parameter $p\lambda$ for $\text{Ba}^+$'s 493~nm is slightly smaller than the corresponding $S_{1/2}$-$P_{1/2}$ transitions in $\text{Ca}^+$, $\text{Sr}^+$ and $\text{Yb}^+$ (up to 10\%. See \cref{table1}), this small disadvantage can be easily offset if the long wavelength of the transition allows for stable operation of high finesse cavities in vacuum.   

In this article we employ two methods to test two potential degradation scenarios: first, oxygen diffusion in vacuum (without any long term exposure to light) and second, laser induced damage, possibly due to deposition of hydrocarbons during prolonged exposure to high intensity radiation on the surface of the mirrors. 
We find that the cavity suffers degradation on prolonged exposure to $493$~nm radiation under ultra-high vacuum (UHV) while, without long exposure, it is stable under UHV conditions for more than sixteen months. 
We discuss potential solutions towards cavity QED with Barium ions or experiments with lasers operating around $493$~nm. Our results are relevant not only for cavity QED with Barium ions but also for experiments involving high-finesse cavities at around $500$~nm. 

The paper is organized as follows.
In \cref{setup}, we describe our experimental setup. We discuss results of tests for vacuum-induced degradation and laser-induced degradation in \cref{vid} and \cref{lid} respectively. 
In \cref{modeandfinesse}, we discuss the dependence of the measured loss on the mode shape that we use to measure the cavity lifetime, indicating the laser-induced nature of the degradation in \cref{lid}. We discuss the recovery process in \cref{sec:recovery}.
 
\section{Experimental setup} 
\label{setup}

\begin{figure}[t!]
    \centering
    \begin{subfigure}[b]{0.68\textwidth}
        \hspace*{-1.5cm} 
        \centering
        \includegraphics[width=\linewidth]{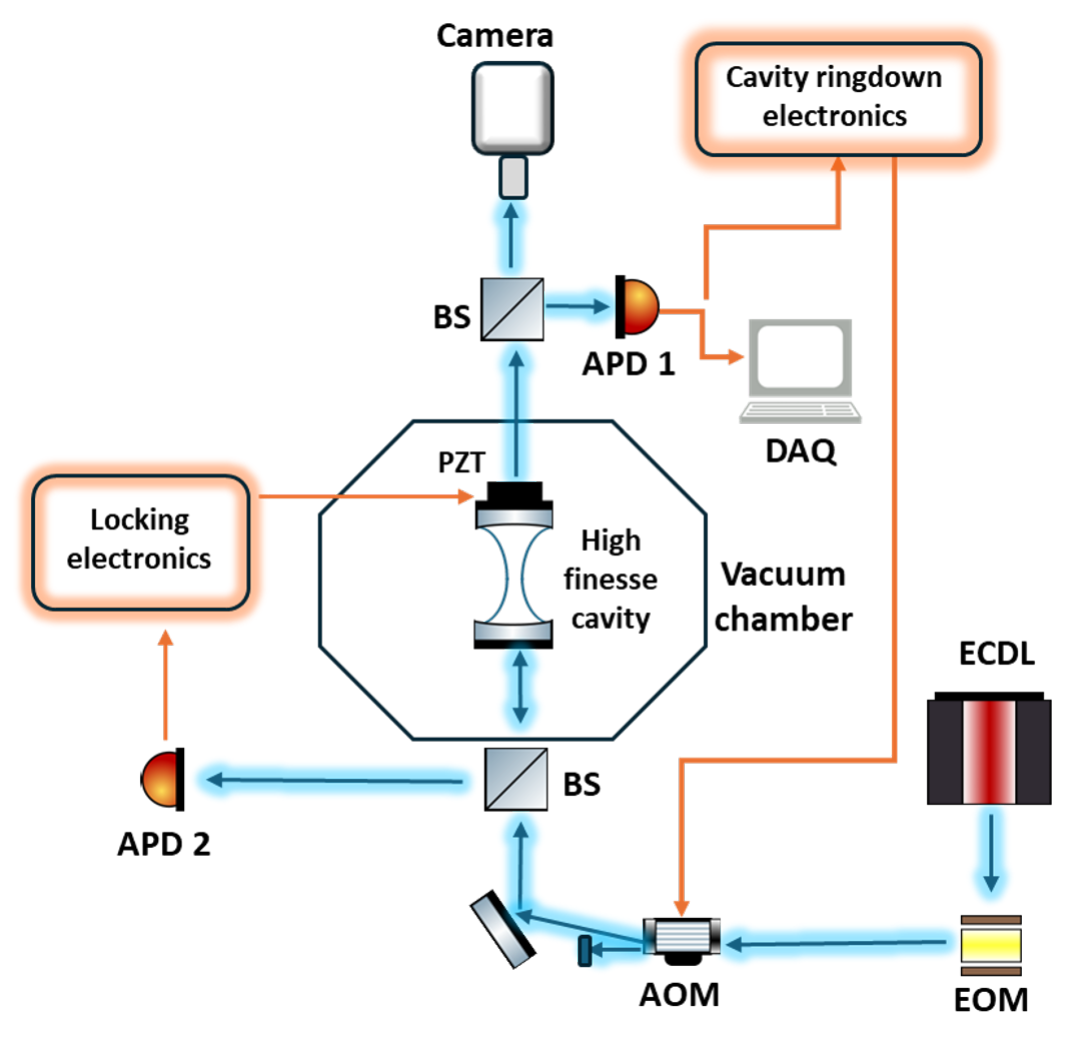}
        \caption{}
        \label{schematic}
    \end{subfigure}
    \begin{subfigure}[b]{0.450\textwidth}
        \includegraphics[width=\linewidth]{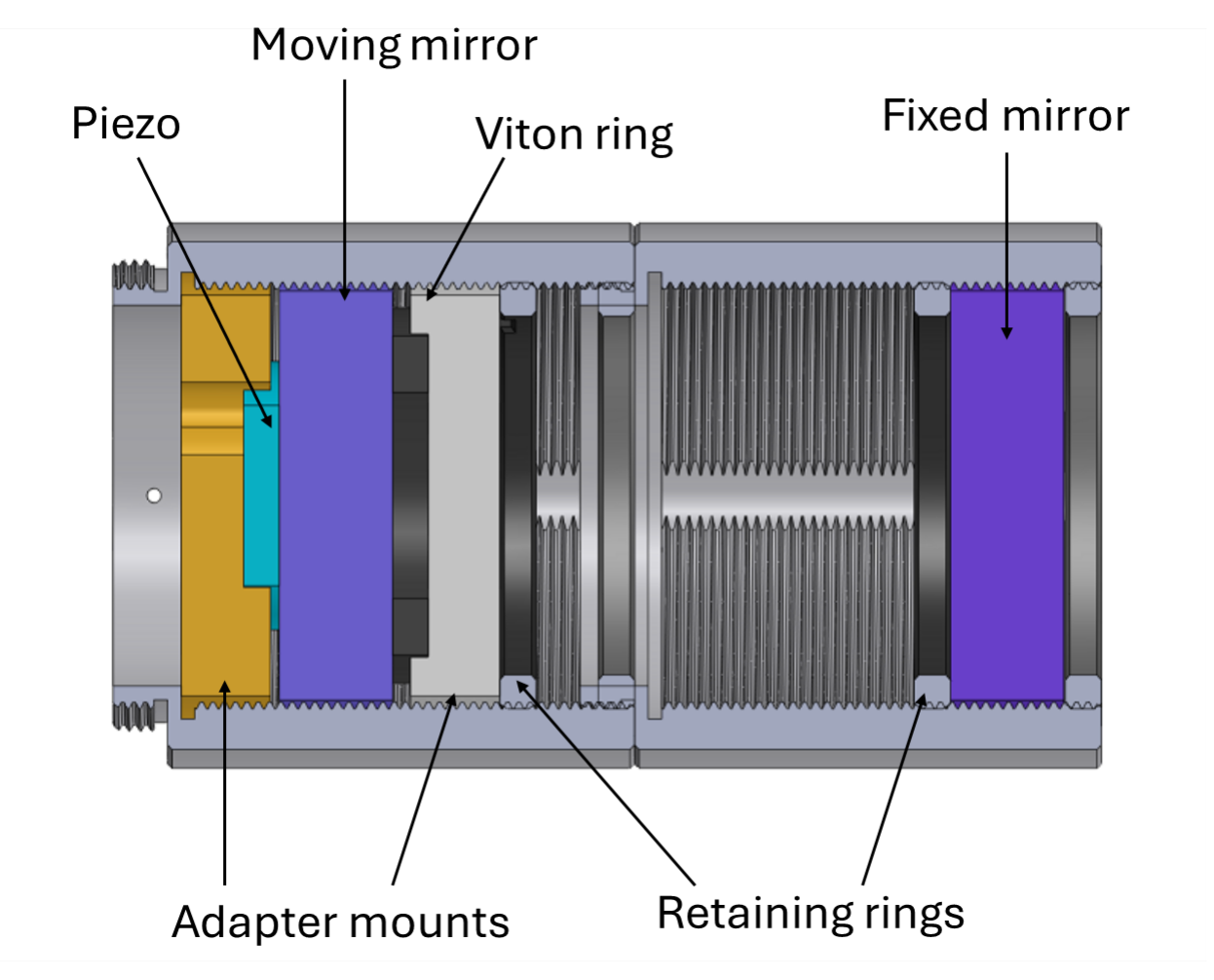}
        \caption{}
        \label{cad}
    \end{subfigure}
    \begin{subfigure}[b]{0.4\textwidth}
        \includegraphics[width=\linewidth]{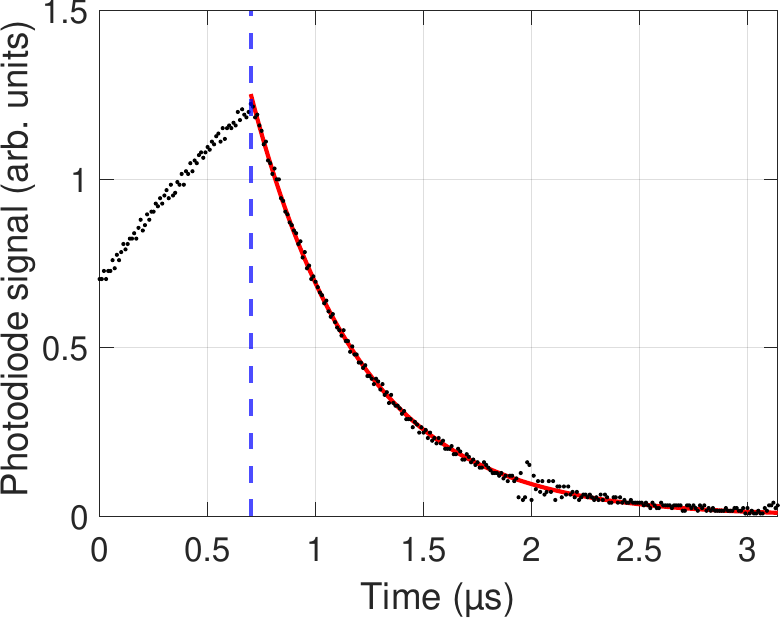}
        \caption{}
        \label{ringdown}
    \end{subfigure}
\caption{(\subref{schematic}) Schematic of the experimental setup. ECDL: External Cavity Diode Laser, APD 1 and 2: avalanche photodiodes detectors for transmitted and reflected signal respectively, BS: beamsplitter, AOM: acousto-optic modulator, EOM: electro-optic modulator, PZT: piezoelectric transducer for scanning the cavity length, DAQ: Data Acquisition system. 
The glowing blue lines indicate optical signal, while the orange line indicates electronic signals. 
(\subref{cad}) Schematic showing a cross sectional view of the cavity assembly.
(\subref{ringdown}) A sample ringdown signal acquired by APD 1 showing exponential decay of cavity transmission after the input beam is turned off. The dashed vertical line indicates the time of extinction of the cavity input signal.} 
\label{figschematic}
\end{figure}

The schematic of the experimental setup is shown in \cref{schematic}.
The test Fabry-Perot cavity was assembled using two mirrors and placed in the vacuum chamber.
The mirrors are coated with stacks of oxide layers (\ce{Ta2O5} and \ce{SiO2}) and are designed to be highly reflective (HR) at $493$~nm (HR493 Laseroptik). 
The cavity is built using a vacuum compatible lens tube with mirrors fitted at the ends with retaining rings (see \cref{cad}). 
One of the mirrors has a ring piezo actuator (Noliac NAC2125) fitted at the back with a compressible Viton ring in front, between the retaining ring and mirror, in order to facilitate scanning of the cavity length.
The length of the cavity is $12.7$ mm.
The mirrors have a radius of curvature of $1000$ mm. 
The entire assembly is mounted on a home-built stage that can be directly fitted to a flange. 

For our $493$~nm lasers to probe the cavity in vacuum, we used different diodes in the external cavity diode laser (ECDL) configuration. 
Initially, we used a NICHIA $493$~nm diode in a home-built ECDL setup. 
Later, we switched to an AR coated diode version of the same diode. 
For experiments requiring long term laser exposure, we used a Toptica TA pro laser at $986$~nm that is frequency doubled to $493$~nm using a fiber coupled second-harmonic generation (SHG) waveguide (NTT Electronics). 
The narrow linewidth of the $986$~nm laser was vital to reduce intensity fluctuations in the locked cavity that were caused by excess phase noise in the NICHIA diodes when operated in the home-built ECDL setup. 
For our experiments, we operate around $493.4$~nm, close to the $S_{1/2}$-$P_{1/2}$ transition of Barium ion.

Throughout the experiment, the pressure of the chamber was maintained at an average of $2\times10^{-8}$ mbar for several months with an ion pump and a non-evaporable getter (NEG) combination pump (SAES NexTORR 200) with pressure fluctuations limited to less than $5\%$. 
We believe that our results will be valid for lower pressures as any previous degradation has been observed at similar or even higher pressures. 
We keep track of the pressure using a Pirani magnetron (Agilent FRG-700) pressure gauge. 

The cavity transmission signal is measured using a sensitive avalanche photodiode (Hamamatsu S12060-10-Si APD). 
We monitor the output mode of the transmitted cavity light using a CCD camera (Thorlabs CS165MU-Zelux) to ensure that the cavity is operating at the TEM\textsubscript{00} mode. 
We are able to excite the TEM\textsubscript{00} mode by overlapping a collimated incoming beam with its reflection from the cavity. 
Based on the cavity parameters, the mode waist in the cavity is calculated to be around $112$~\unit{\micro\meter}. 

Cavity ringdown is a standard method for measuring the lifetime and loss of optical cavities \cite{okeefe1988cavity}. 
The signal is processed by a home-built comparator circuit (AD 8561) operating with finite hysteresis. 
The comparator triggers "high" when a threshold voltage is reached and turns to "low" at a lower voltage level. 
The cavity signal input is derived from the first order diffraction from an acousto optic modulator (AOM). 
During a slow cavity scan, once the comparator detects a threshold cavity signal, a TTL trigger turns off the RF signal to the AOM. 
Following a delay of about $600$~ns after the signal turns off (consistent with the speed of the acoustic signal in the crystal), it takes approximately $20$ ns (as measured by the 90\%-10\% fall time) for extinction of the input signal. 
This is fast compared to the cavity lifetime. 
After extinction of the cavity input, the electric field already built up inside the cavity leaks over a characteristic timescale given by the cavity lifetime. 
The hysteresis in the comparator circuit prevents the AOM RF signal from turning back on after the output signal drops below the triggering threshold, allowing a period of exponential decay of the cavity field. 
A sample ringdown signal with a fit is shown in \cref{ringdown}.
The photodiode signal $P_\text{D}(t)$ after extinction of the input beam (denoted as the dashed vertical line in \cref{ringdown}) is numerically fit to an exponential function:
\begin{equation}
    P_\text{D} (t) \propto e^{-\frac{t}{\tau}}. 
\label{finessedecaytimeeq1}
\end{equation}
The cavity loss $L$ is related to the cavity lifetime $\tau$ as
\begin{align}
    L + 2T = \frac{2l}{c\tau},
\label{eq:tau}
\end{align}
where $T$ is transmission efficiency of the mirror and $c$ is the speed of light.
Using \cref{eq:tau}, the cavity loss can be obtained from the measurement of the cavity lifetime. 
On the other hand, the cavity finesse is related to the cavity lifetime as
\begin{equation}
    F= \frac{\pi c\tau}{l}.
\label{eq:finesse-decaytime}
\end{equation}
At the start of the experiment, we measure the cavity lifetime to be $367\pm3$~ns, which translates to a cavity loss of $31\pm2$ ppm while using $T = 100$ ppm specified by the mirror manufacturer.
It also corresponds to a finesse of $27200\pm200$. 
In the following sections, numerical results are reported in terms of measured cavity loss, which is related to the cavity finesse by \cref{eq:finesse-decaytime}. 
The terms are sometimes used interchangeably in the paper.

\section{Experiment 1: Test for surface oxygen depletion or other vacuum-induced damage} 
\label{vid}

In this section, we explore the possibility of degradation of the cavity due to a prolonged stay in vacuum. 
This may be caused by oxygen depletion (or diffusion) from the surface of the mirrors, leading to the formation of absorbing colour centers as discussed in \cite{Gangloff2015}. 
Vacuum induced losses are also discussed in \cite{Gallego} and \cite{Brandstatter2013}. 
To keep track of any degradation, we measure the cavity loss using the cavity ringdown technique. 
The method is very sensitive to variations in reflectivity and surface losses that are likely to arise from any damage to the oxide coating.

\begin{figure}[h]
    \centering
    \includegraphics[width=.8\textwidth]{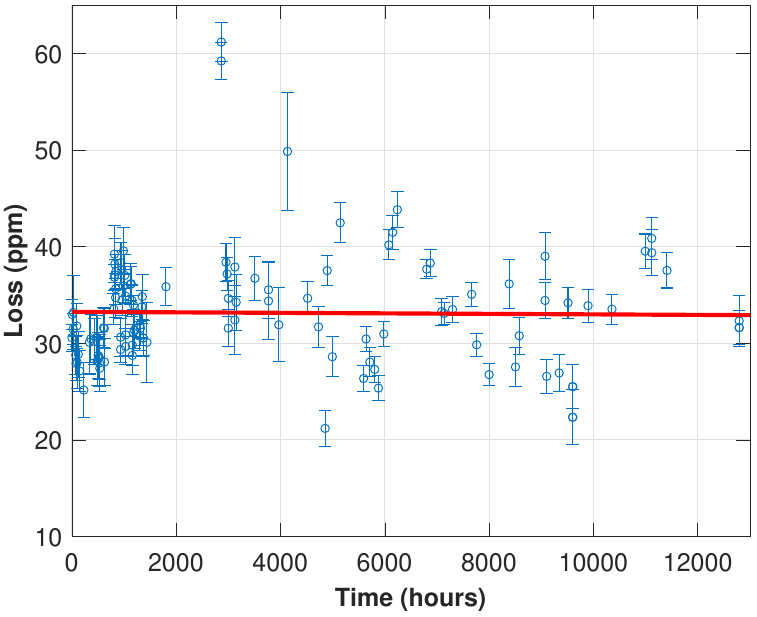}
    \caption{Cavity loss measured in vacuum (blue open circles) over several months. 
    The error bars are standard deviations of hundred measurements. 
    Also shown in figure is the linear fit (red line) to the data.}
    \label{fig1}
\end{figure}

The results are presented in \cref{fig1}. 
The data shown are compiled over a period of more than a year. 
After several months, we observe very little change in cavity loss. 
There are small fluctuations in finesse that we believe are related to changes in alignment of the probe beam into the cavity. 
Before installation in the chamber, the mirrors had accumulated some dust and had to be cleaned. 
However, the cleaning was imperfect, and the mirrors ended up retaining some patches of dirt, resulting in slightly varying reflectivity across the surface. 
This also results in some variations in cavity loss that show up in our data from time to time as the alignment of the input beam drifts. 
We estimate the circulating power $P_\text{c}$ inside the cavity by 
\begin{equation}
    P_{\text{c}}=\frac{\eta_m T}{\qty(L+2T)^2}P_{\text{in}}.
    \label{eq:circulatingpower}
\end{equation}
Here $\eta_m$ is the mode matching efficiency of the input beam to the TEM\textsubscript{00} mode. 
The term $P_{\text{in}}$ is the input power at the cavity mirror, which is calculated by measuring the power outside the vacuum chamber and taking into account optical losses at the chamber window. 
During the experiments, the only exposure to $493$~nm radiation was during the ringdown measurements. 
The peak circulating power in the cavity is less than $400$~mW and the total duration of exposure during the ringdown measurement is on the order of micro-seconds. 

We do not observe any evidence of degradation of the oxide coating due to the vacuum environment. 
From our data, a linear fit gives the rate of change in cavity loss to be $(0.4\pm1.5)\times10^{-4}$ ppm/hour, consistent with zero. 
Our results are also consistent with \cite{Gangloff2015} where vacuum induced degradation due to oxygen diffusion has been reported to be reduced when the top layer is silica. 
However, a silica top layer has also been shown to be ineffective in preventing damage in other experiments \cite{Ballance2017}, implying the possibility of multiple degradation mechanisms.

\section{Experiment 2: Tests for laser-induced damage} 
\label{lid}

\subsection{Experiment with prolonged exposure of the cavity to resonant light} 
\label{sec:prolonged_exposure}

\begin{figure}[th]
    \centering
    \includegraphics[width=.75\textwidth]{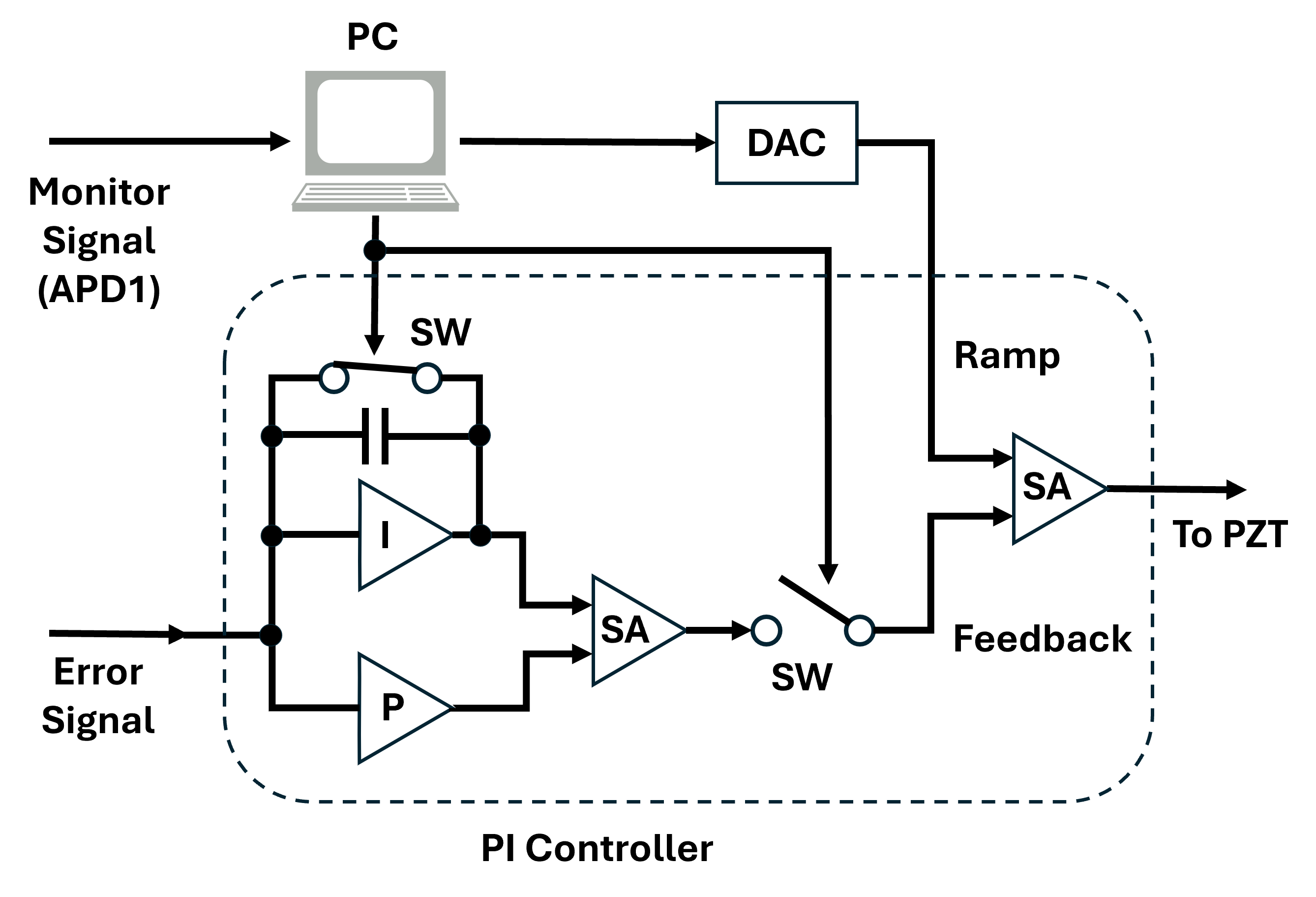}
    \caption{Schematic of the electronic control system for cavity relocking.
    Enclosed with a dashed line is the PI controller implemented on a PCB. 
    DAC: digital-to-analog converter, P: P-gain, I: I-gain, SA: summing amplifier, SW: analog switch.}
    \label{fig:relock}
\end{figure}

\begin{figure}
    \hspace*{-1.0cm} 
    \begin{subfigure}{0.52\textwidth}
        \includegraphics[width=\textwidth]{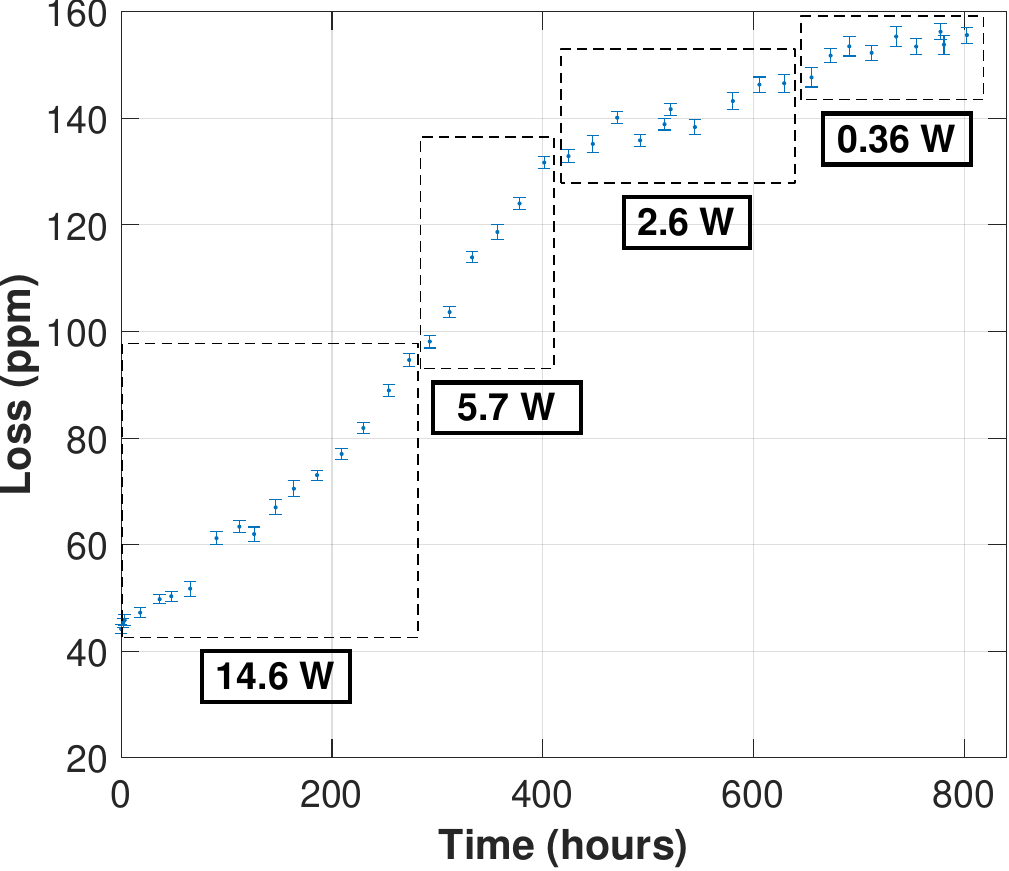}
        \caption{}
        \label{cavityloss-run1}
    \end{subfigure}
    \hfill
    \begin{subfigure}{0.52\textwidth}
        \includegraphics[width=\textwidth]{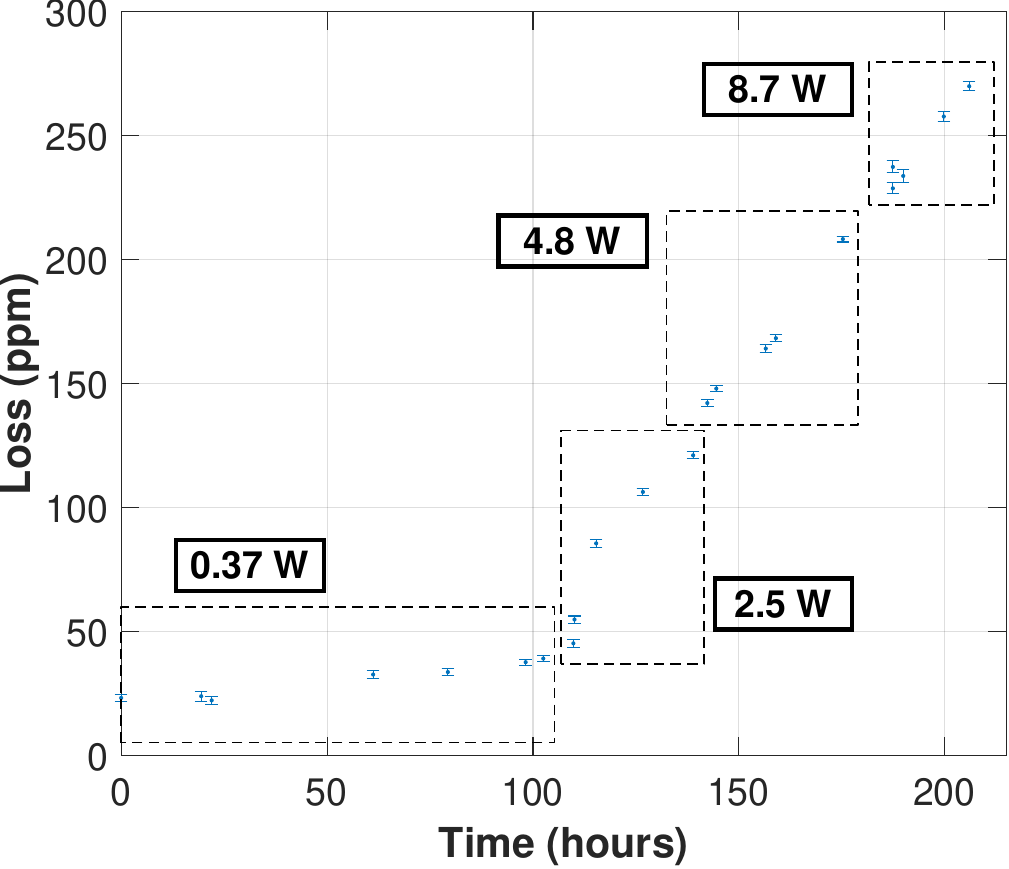}
        \caption{}
        \label{cavityloss-run2}
    \end{subfigure}
    \caption{Cavity loss measured as a function of time for two different runs, (a) and (b), detailed in the main text. 
	Loss values inside the dashed boxes are measured at mean circulating powers denoted in the solid boxes.}
    \label{fig2}
\end{figure}

In this experiment, we measure the cavity loss after prolonged and continuous exposure to resonant light to test for laser induced degradation. 
Light at ultraviolet wavelengths is known to cause damage to optics in vacuum \cite{wagner2014laser}. 
Laser-induced damage to mirror surfaces caused by the deposition of hydrocarbon material was tested under laboratory conditions with a pulsed $355$~nm laser in \cite{wagner2014laser}. 
The coating in that experiment indicated signs of hydrocarbon deposition on the spot exposed to laser. 
The evidence for the damage being laser-induced was the shape and location of the deposit coinciding with the laser spot on the mirror. 

\begin{figure}[h]
    \centering
    \includegraphics[width=.8\textwidth]{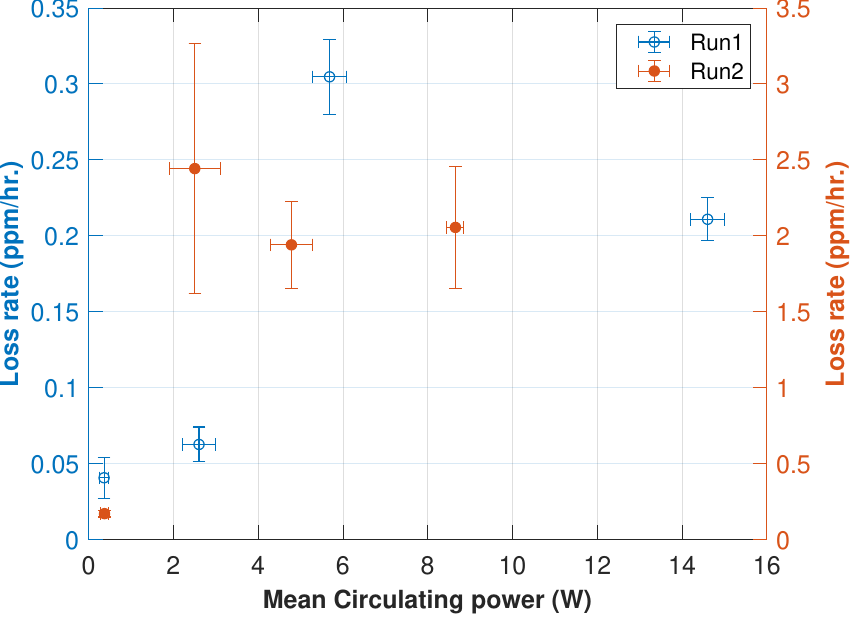}
    \caption{Loss rates (ppm/hr.) as a function of mean circulating power for the two different runs. The vertical error bars originate from the errors in fitting the loss data with a linear function. The horizontal error bars are the errors in estimating the circulating optical powers in the cavity due to the continuous degradation of the cavity between the loss measurements (see the main text). }
    \label{lossincreaserate}
\end{figure}

For this experiment, we use the frequency doubled $986$ nm laser as the $493$ nm laser source to inject optical powers into the cavity. 
Because of the high finesse of the cavity, the laser's narrow linewidth of $\sim 200$ kHz is critical to have small intensity fluctuations in the locked cavity. 
We lock the cavity to the laser at the resonance of TEM\textsubscript{00} mode using the Pound Drever Hall (PDH) technique \cite{drever1983laser}. 
With the cavity locked at resonance to the laser, there is a large enhancement of laser power inside the cavity, leading to high circulating powers. 
The circulating power inside the cavity is calculated using \cref{eq:circulatingpower}. 
The laser is free running, as we did not notice any strong drifts in wavelength or severe mode hops.
To ensure cavity locking for a prolonged time, we implemented an electronic control system that automatically relocks the cavity when it becomes unlocked.
As illustrated in \cref{fig:relock}, the cavity is locked by feeding back the error signal via a PI controller while the optical transmission through the cavity is monitored with a PC. 
The PI controller is a home-made analog electronics that incorporates analog switches (see \cref{fig:relock}).
Conditioned on an event where the transmission goes below a certain threshold signalling unlocking of the cavity, the PC initiates the relocking process: TTL signals are sent to the analog switches which reset the integral gain of the PI controller and disconnect the output from the PZT, respectively. 
At the same time, a voltage ramp controlled by the PC is applied to the PZT to scan the cavity length.
Once the resonance is found and the transmission goes above the threshold, the PC stops the ramp and sends TTLs in the reverse logic to engage the lock again.    
We gather data for several days, measuring the cavity loss at regular intervals at various circulating powers while ensuring that we stay well below the damage threshold of the mirrors (reported by the manufacturing company to be 10 kW/cm).  

We conducted two separate runs of measurements.
In the first run we started with a high circulating power and subsequently lowered the power in multiple steps.
An increase in cavity loss was observed within the first few days (see \cref{cavityloss-run1}).
The cavity loss was measured regularly, at which point the continuous exposure was stopped for a period between 20 minutes and half an hour.
We tried to maintain a steady circulating power, updating the input power to compensate for the increase in cavity loss. However, the adjustment was made only after the ringdown measurement. On average, this was performed at intervals of around 21 hours. Hence, the circulating power gradually decreased over this duration as the cavity degraded. 
After measuring the cavity loss, we lock the cavity to the laser again, update the input power, and continue the experiment, exposing the cavity mirrors to the radiation until the next measurement. 
We observed that the loss continued to increase at a steady rate, slowing down only when the circulating power was lowered. 
Cavity loss was found to increase from 30 ppm to more than 130 ppm over a total of 800 hours of exposure. 
At a certain circulating power, we quantify the rate of increase in loss by a linear fit to the change in cavity loss as a function of the duration of exposure of the cavity to radiation. 
The measured cavity loss at four different circulating powers are shown in \cref{cavityloss-run1}.

After completing the first run of the measurements shown in \cref{cavityloss-run1}, we vented the chamber to atmospheric pressure to test the recovery process in the presence of oxygen. 
We will discuss the recovery process in detail in \cref{sec:recovery}. 
For now, we mention that the increase in losses was reversed and the cavity loss recovered to its original value. 
After recovery, we started the second run of measurements after reaching the pressure of $2\times10^{-8}$ mbar, the same as in the first run.
Note that the chamber had to be opened for some technical issues before the pump down. However, any potential contamination is minimal and the cavity was always in a closed tube.
In the second run, we wanted to test the possibility of a high circulating power triggering a degradation mechanism that after being triggered, continues at lower powers. 
To do so, we started with low circulating powers and then increased the power. 
We started with a circulating power of around $370$ mW. 
This is similar to the circulating power for the last few data points in the first run (\cref{cavityloss-run1}), and is also similar to what we have used during cavity loss measurements in the \cref{vid}. 
We then moved to higher circulating powers for subsequent measurements.  

We find a much higher rate of increase in loss during the second run as shown in \cref{cavityloss-run2}. 
Even at the lowest circulating power, the rate of increase in loss was found to be higher than what was observed at much higher circulating powers during the previous run. 
The comparison is shown in \cref{lossincreaserate} where we have used a simple linear fit to quantify the loss rate. The discrepancy in the rates of increase in loss values indicates very high sensitivity to even small changes in the local environment around the cavity.
We do not observe a threshold circulating power that triggers the loss mechanism. However, the results show that short exposure at low circulating power is unlikely to cause noticeable degradation. 
Damage occurs on timescales of a few hours and depends on the circulating power. 
Therefore, applications involving low circulating powers are possible at this wavelength. 
Cavity QED experiments using the Barium ion's $S$-$P$ transition at a single or few photon level can be carried out. 

\subsection{Cavity spatial modes and losses} 
\label{modeandfinesse}

Here we report loss measurements using different transverse spatial modes in a degraded cavity. 
For our long-term measurements presented in \cref{sec:prolonged_exposure}, we locked the cavity using the fundamental TEM\textsubscript{00} mode.
However, we find that measurements of cavity lifetimes that are performed using the TEM\textsubscript{01} and TEM\textsubscript{11} modes show lower cavity losses. 
We believe that this is because the spatial distribution of the damage incurred on the mirror follows that of  the TEM\textsubscript{00} mode used for locking the cavity, and hence the losses experienced by these higher spatial modes are relatively smaller. 

\begin{figure}[!t]
    \centering
    \includegraphics[width=.8\textwidth]{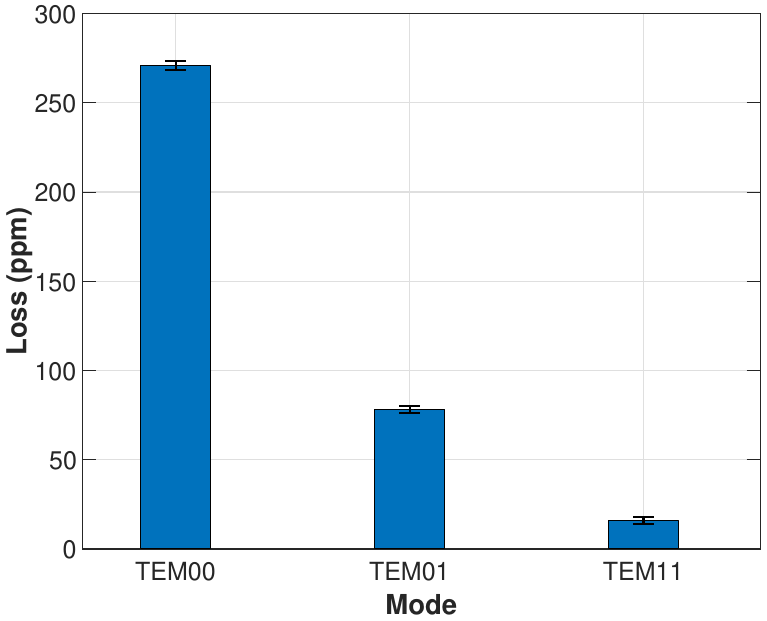}
    \caption{Variation of the cavity loss with the transverse spatial modes of the degraded cavity.} 
    \label{fig:modes-vs-loss}
\end{figure}

The cavity losses were measured with the ringdown method mentioned in the previous section. 
In order to excite the higher order modes, alignment into the cavity was slightly altered, without moving away too far from the original spot. 
After the realignment, all the measurements shown as the blue bars in \cref{fig:modes-vs-loss} were carried out by scanning the cavity length to excite different modes without moving the beam. 
The change in finesse of the TEM\textsubscript{00} mode due to this shift from the original spot was around $3\%$ within the error of the measurement, indicating that the damaged section of the mirror was still in overlap with the TEM\textsubscript{00} mode. 
The figure shows that higher order modes experience lower losses as they have a smaller overlap with the region of damage induced by the TEM\textsubscript{00} mode. 
This is evidence of the damage being laser induced and being localized around the irradiated spot.

\subsection{Potential role of thermally activated processes}

At high circulating powers, local heating of the mirrors at the laser irradiation spot can potentially accelerate the oxygen depletion process, which has been argued to be temperature dependent \cite{Gangloff2015}. 
This can lead to the formation of color centers, which causes an increase in cavity loss. In this Arrhenius-type process the rate of increase in loss has a temperature dependence given by $\exp(-U/k_\text{B}T)$, where $U$ is the activation energy of the process, $T$, the temperature in Kelvin and $k_\text{B}$ is the Boltzmann constant. 
High-temperature baking and annealing have been shown to degrade mirrors by oxygen depletion \cite{Brandstatter2013}.

Such a mechanism would also be consistent with the damage being laser induced, which we concluded from our results in \cref{modeandfinesse}. 
However, our experiments do not support this hypothesis. 
First, for similar circulating intensities, we observed higher loss rates for the second run that was carried out after complete recovery (see \cref{lossincreaserate}). 
For a thermally activated process, this should not be the case. 
On the other hand, rates of laser induced deposition may vary depending on the environment in the chamber, and the possible increase in contamination in the chamber in the second run better explains the drastic increase in loss/ppm values. 
Second, we have the fact that the top layer of both our mirrors is silica that, as discussed in \cite{Brandstatter2013} has a much higher activation energy for the oxygen depletion process than tantalum (V) oxide. 
Third, in both runs, the rate of increase in loss does not show a steady increase with increasing circulating power (see \cref{lossincreaserate}), though the temperature of the irradiated spot should increase with increasing circulating powers. 
In addition, if the irradiation merely heats the surface and accelerates the rate of increase in loss, we should have observed a finite rate of increase in loss even at room temperature and without prolonged exposure to radiation. 
As discussed in \cref{vid}, we do not find evidence for this. Thus, it is very unlikely that thermally induced processes are the primary cause of cavity degradation. 
Although it can potentially play a role in combination with laser induced deposition, we expect its contribution to be small. 

\section{Recovery} 
\label{sec:recovery}

\begin{figure}[h]
 \centering
    \includegraphics[width=.78\textwidth]{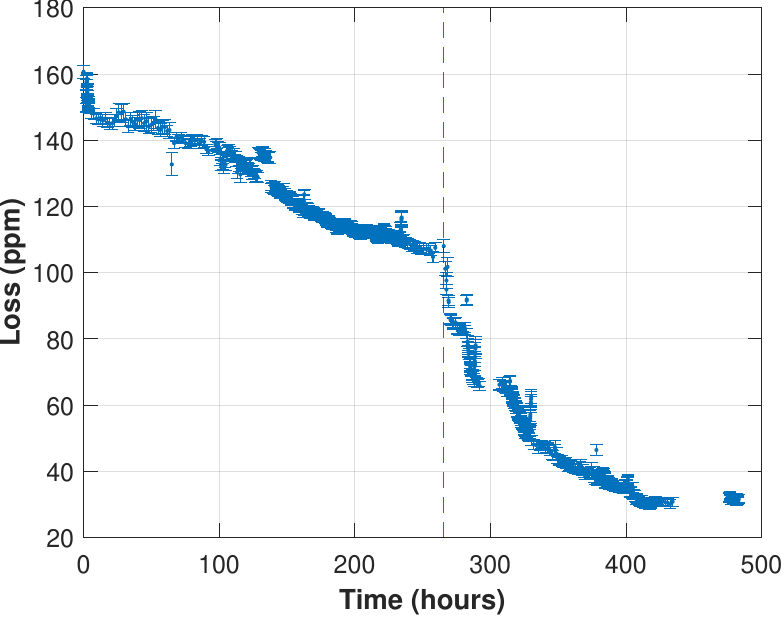}
    \label{Recovery: Cavity Loss vs Time exposed}
    \caption{ Cavity loss as a function of time during the recovery phase.
	The dashed line indicates the time when the cavity was irradiated with the 397~nm beam, leading to a clear increase in the rate of recovery. 
	The first data point marks the time of chamber being exposed to atmospheric air. 
	The final data-point indicates a complete recovery.}
    \label{figrecovery}
\end{figure}

Here, we describe the process by which the increase in loss can be reversed and the finesse recovered. 
The recovery of finesse on exposure to oxygen has been reported in \cite{Ballance2017,Gangloff2015, Gallego}. 
First, we vented the chamber while purging it with nitrogen, which did not lead to any change in the measured cavity loss. 
After a day, we slowly introduced atmospheric air and noticed a rapid initial recovery, presumably aided by atmospheric oxygen. 
This was soon followed by a slow but steady recovery. 
We plot the cavity loss as a function of time starting from the introduction of atmospheric air in \cref{figrecovery}.
The initial rapid recovery is seen in the early section of the plot.

In \cite{Gangloff2015} and \cite{Gallego} the recovery was assisted by exposure to UV radiation at $422$ nm and $405$ nm respectively. 
In our case, we observe an improved rate of recovery using radiation at $397$ nm. 
The cavity mirrors have low reflectivity at $397$ nm and the light is mostly transmitted. 
We used $5$ mW of the $397$ nm laser at a spot size of $292$ \unit{\micro\meter}, fully overlapping with the damaged spot.
As indicated in \cref{figrecovery} by a dashed vertical line, there is a clear increase in the rate of recovery on the introduction of $397$ nm laser.

Like the process of damage, the mechanisms of recovery are not very well understood either. In \cite{Gangloff2015}, a reversal of the oxygen diffusion process is suggested. In our case, however, the damage is clearly laser-induced, unlike oxygen diffusion in vacuum. However, as shown in \cref{figrecovery}, recovery is seen to be accelerated by exposure to radiation at $397$ nm. 

The combination of ozone and oxygen free radicals can break down hydrocarbon deposits on surfaces \cite{vig1985uv}. 
However, it is unclear whether there is enough energy in $397$ nm photons to form ozone in the first place. 
The mirrors ended up showing slightly higher finesse than that measured at the beginning, indicating the effectiveness of the $397$~nm laser and oxygen in reversing damage to the mirrors. 
Note that there seem to be variations in the values of cavity loss in the data. This is due to the small variations in alignment of the $493$~nm beam used for ringdown measurements. This results in the cavity loss being measured at slightly different spots.
Though not indicated in the figures, we verified that partial overlap of the spot with $397$~nm laser reduces the rate of recovery. Therefore, a higher power is likely to further accelerate recovery.

\section{Discussion and Conclusion:}
We show that a cavity made of oxide-coated mirrors, designed to operate at $493$~nm, shows no degradation in vacuum and on exposure to resonant radiation at low circulating powers. 
We observe damage on prolonged exposure at higher circulating powers. 
We establish the laser induced nature of damage by measuring cavity loss using higher order modes. 
In addition, we rule out the role of thermally activated processes being responsible for degradation.

As in the case of experiments at deep-UV, the mechanisms are not well understood. 
The damage is established to be laser induced although it is prominent at high circulating powers over a time scale of a few days. 
Measurements of cavity decay time can be used to detect variations in cavity loss at the ppm level, making the experiment suitable to probe such changes \cite{Rempe1992}. 
The silica top layer, though robust in vacuum over several months and on short exposures at low powers, was not enough to prevent laser induced damage at higher powers. 
A complete recovery of finesse required exposure to oxygen and was accelerated by radiation at $397$ nm. 
On the basis of our results, we believe that laser-induced deposition is the most likely explanation for the observed results. 

These observations may create issues in the integration of ion traps within optical cavities using Ba\textsuperscript{+} ions. 
Our results show that conducting cavity QED experiments at low photon numbers is possible, as the cavity does not degrade in the absence of strong irradiation. 
However, locking the cavity at the $493$ nm  transition can cause problems. 
Circumventing the problem by using other methods of locking the cavity may be necessary. 
For instance, with a coating of mirrors that is strongly reflecting at two wavelengths, a longer wavelength may be used as a reference laser to lock the cavity to resonance. 
Our results at high intra-cavity powers are relevant for experiments that seek to use the cavity standing wave as optical traps for atoms or ions \cite{Home}. 
With a cleaner environment around the cavity and lower circulating powers, we expect a lower damage. 
The use of calcium fluoride coatings after conditioning with ultraviolet radiation has shown stability at high circulating powers, though it is not clear whether it is robust against laser-induced deposition over long timescales \cite{burkley2021stable}. 

During the baking and initial pumpdown of a vacuum system, heavier hydrocarbons can be difficult to remove. 
Using the ozone cleaning method, heavier hydrocarbons can be broken into lighter ones that are easier to pump out \cite{wanzenboeck2010novel}. 
This can create a cleaner environment for the high finesse cavities and inhibit laser induced degradation in vacuum. 
The removal of all hydrocarbon materials is also a promising option\cite{Schmitz2019}, though it may not be feasible for many applications. 
A deeper understanding of the mechanisms of damage as well as recovery may also be of interest for space-based applications where continuous exposure to visible and UV radiation in vacuum can shorten the lifetime of optics and in engineering ultrastable optical frequency references, which can benefit from a vacuum environment.

\section{Back matter}

\begin{backmatter}
\bmsection{Funding}
This work was supported by JST Moonshot R\&D (Grant Number JPMJMS2063).

\bmsection{Acknowledgment}
The authors acknowledge help from the OIST machine shop and for machining and deanodizing parts for in-vacuo setup, Shigeo Sugitani for help with electronic components and Makoto Endo for his contribution in the early stage of the experiment.

\bmsection{Disclosures}
\medskip
\noindent The authors declare no conflicts of interest.

\bmsection{Data Availability Statement}
Data available upon request.
\end{backmatter}

\bibliography{wg2_cavity}

\end{document}